\documentclass[fleqn,usenatbib]{mnras}

\usepackage[T1]{fontenc}

\DeclareRobustCommand{\VAN}[3]{#2}
\let\VANthebibliography\thebibliography
\def\thebibliography{\DeclareRobustCommand{\VAN}[3]{##3}\VANthebibliography}

\usepackage{graphicx}	
\usepackage{amsmath}	
\usepackage{amssymb}	

\newcommand{\kms}{\,km\,s$^{-1}$}
\newcommand{\kmss}{km s$^{-1}$}
\newcommand{\km}{km s$^{-1}\;$}

\newcommand{\vlsr}{V$_{\rm LSR}$}

\newcommand{\lsun}{\mbox{L$_{\sun}$}}

\newcommand{\ho}{H$_{2}$O$\;$}

\newcommand{\hco}{HCO$^{+}\;$}
\newcommand{\hcn}{HCN$\;$}
\newcommand{\mb}{mJy beam$^{-1}$}
\newcommand{\nhd}{$n_{\rm H_2}$}

\newcommand{\nhh}{$N$$_{\rm H}$}

\title[Off-nuclear H$_2$O maser in NGC\,1068]{Off-nuclear H$_2$O maser and dense molecular gas in NGC\,1068}

\author[Y. Hagiwara]{
Yoshiaki Hagiwara,$^{1,2,3}$\thanks{E-mail: yhagiwara@toyo.jp (YH)}
Willem  A. Baan,$^{2,4}$ Masatoshi Imanishi,$^{5,6,7}$ and Philip Diamond$^{8}$  
\vspace{5mm}
\\
$^{1}$Natural Science Laboratory, Toyo University, 5-28-20, Hakusan, Bunkyo-ku, Tokyo 112-8606 Japan\\
$^{2}$Netherlands Institute for Radio Astronomy ASTRON,  7991 PD Dwingeloo, the Netherlands\\
$^{3}$Joint Institute for VLBI ERIC, 7991 PD Dwingeloo, the Netherlands\\
$^{4}$XinJiang Astronomical Observatory, Chinese Academy of Sciences, Urumqi, PR China \\
$^{5}$National Astronomical Observatory of Japan, National Institutes of Natural Sciences (NINS) 2-21-1 Osawa, Mitaka, Tokyo 181-8588, Japan  \\ 
$^{6}$Department of Astronomy, School of Science, The Graduate University for Advanced Studies, SOKENDAI \\ 
2-21-1 Osawa, Mitaka, Tokyo 181-8588 Japan \\ 
$^{7}$Toyo University, 5-28-20, Hakusan, Bunkyo-ku, Tokyo 112-8606 Japan\\
$^{8}$SKA Observatory Jodrell Bank, Lower Withington, Macclesfield, SK11 9FT, UK}
\date{Accepted XXX. Received YYY; in original form ZZZ}
\pubyear{2023}
\begin{document}
\label{firstpage}
\pagerange{\pageref{firstpage}--\pageref{lastpage}}
\maketitle

\begin{abstract}
The results of high-resolution spectral-line observations of dense molecular gas are presented towards the nuclear region of the type 2 Seyfert galaxy NGC\,1068. 
 MERLIN observations of the 22 GHz H$_2$O maser were made for imaging the known off-nuclear maser emission at radio jet component located about 0.3" north-east of the radio nucleus in the galaxy. High angular resolution ALMA observations have spatially resolved the molecular gas emissions of HCN and HCO$^{+}$ in this region. 
The off-nuclear maser spots are found to nearly overlap with a ring-like molecular gas structure and are tracing an evolving shock-like structure, which appears to be energized by interaction between the radio jet and circumnuclear medium. The scenario of the dynamic jet-ISM interaction is further supported by a systematic shift of the centroid velocities of 
the off-nuclear maser features over a period of 35 years. The enhanced integrated flux ratios of the HCN to HCO$^{+}$ line emission features at component C suggest a kinetic temperature T$_{k}$\ga 300K and an H$_2$ density of \ga10$^6$ cm$^{-3}$, which are conditions where water masers may be formed.
The diagnostics of the masering action in this jet-ISM interaction region is exemplary for galaxies hosting off-nuclear \ho maser emission.
\end{abstract}

%
\section{Introduction}
Highly luminous extragalactic \ho maser in the transition of $6_{16}-5_{23}$ (rest frequency: 22.23508 GHz) is a useful tracer of the structure and kinematics of dense molecular gas in interaction regions in galaxies and in clouds on scales of $<$ $\sim$1 pc from a central engine of active galactic nuclei (AGN). 
Some fraction of the known \ho megamasers exhibit spectra implying evidence for emission sub-parsec clouds embedded in an edge-on rotating disc surrounding super massive black holes using Very Long Baseline Interferometry (VLBI) at milliarcsecond (mas) angular resolution \citep[e.g.,][]{nak93, has94, miyo95, link03,baa22}. Some of the megamaser systems were studied by the Megamaser Cosmology Project (MCP)\footnote{\url{https://safe.nrao.edu/wiki/bin/view/Main/MegamaserCosmologyProject}, 2020-06-15, Jim Braatz}, aiming the independent and high precision measurement of the Hubble constant at a 2-3 percent accuracy by measuring the megamasers hosting the "clean masering disk" systems, as in NGC4258, in the Hubble flow with VLBI mapping and spectral monitoring. \citep[e.g.,][]{rei09,gao16,pes20}.

Many of the \ho megamasers are associated with the radio continuum emission structures resulting from nuclear radio jets, jet-cloud interaction regions or outflows such as TXFS 2226-184, NGC 1052, Mrk 348, and Centaurus A \citep[e.g.,][]{koe95,cla98,pec03,kame05,ott13,sur20}. 
These four objects stand out from the other megamasers on the basis of their line profiles. 
The spectral line profiles of non-nuclear \ho memgamasers consist of single, broad and, relatively smooth emission profiles (FWHM$\sim$100--200 \kmss) and are different from profiles found in nuclear masers that consist of narrow components or "high-velocity" components with line-widths of a few to $\sim$10 \kms spread over much wider velocity ranges of $\sim$100-1000 \kmss. These non-nuclear masers arise through the shock excitation of ambient gas in circumunuclear regions of galaxies. 
The masers are associated with the radio continuum emission that believed to be coincident with the nucleus.



Sub-millimetre (sub-mm) telescopes such as Atacama Large millimetre/sub-millimetre Array (ALMA) have opened the possibility to make imaging studies of (sub-)millimetre masers such as \ho maser in the transitions of 3$_{13}$$-$2$_{20}$ (rest frequency: 183.308\,GHz) and 10$_{29}$$-$9$_{36}$ (rest frequency: 321.226\,GHz) at high angular resolution to $\sim$ 20 milliarcsecond (mas), which may serve to diagnose the masering processes in \ho megamaser sources \citep[e.g.,][]{liz05,hagi13,hagi21}.
The detection of thermally excited molecules such as HCN and HCO$^{+}$ are also reported in these megamaser host galaxies at higher transitions at sub-millimetre wavelengths. 

%

%

The galaxy NGC\,1068 is a prototypical type 2 Seyfert galaxy hosting an active nucleus obscured by a torus in which broad optical emission lines are seen in polarized optical spectra \citep[e.g.][]{ant85}. 
The galaxy has been investigated extensively at all wavelengths, revealing that various species of molecular gas are abundant in the circumnuclear region, and spiral arms in the galaxy and abundant dense molecular gases related to star-forming activity are in the starburst ring \citep[e.g.][]{tak14,gar19,san22}.
NGC\,1068 hosts a nearly one-sided radio jet that may be qualified as a Compact Symmetric Object and luminous and variable 22 GHz \ho maser emission has been detected to be superposed on the radio core-jet structure at lower frequencies \citep[e.g.][]{gal01,gal04}.  
Sub-mm \ho maser action in the 183 and 321 GHz transitions towards the centre of the galaxy has not yet been detected using ALMA \citep{hagi16, pes16, pes23}.
Very long baseline interferometry (VLBI) observations have revealed that the prominent 22 GHz \ho maser features in NGC\,1068 covering a frequency range from \vlsr = 800 -- 1500 \kms are distributed along a linear $\sim$1 pc region and originated in an edge-on ($i$ $>$ 80$^{\circ}$) molecular gas disc in the nuclear region of the galaxy \citep[e.g.][]{linc96,gal01}. 

Recent sub-mm mapping observations of thermally-excited molecular emission lines in the galaxy using ALMA revealed that a parsec-scale torus structure, that surrounds the nuclear region, does not have the simple axisymmetric structure expected in the unified theory \citep[e.g.][]{imp19, ima18, ima20}.


\citet{gal96} reported the presence of \ho maser towards radio jet component C (hereafter the ''off-nuclear maser''), which is located about 0.3 arcsecond ($\sim$ 20 pc) north-east from the component S1. The off-nuclear maser is peaked at \vlsr =  984.1 \kms  \citep{gal01}, which is blueshifted $\sim$140 \kms with respect to the systemic velocity of \vlsr = 1123 $\pm$ 4 \kms, estimated from CO emission \citep{nak95}.
This off-nuclear maser region traces shocked dense molecular gas coinciding with the radio jet component C, where the jet interacts with the local interstellar medium (ISM) in the galaxy \citep{gal96}. Off-nuclear masers will thus provide additional diagnostics of the circumnuclear environment of the AGN.


Following-up the earlier studies using VLBI for clarifying the origin of the off-nuclear \ho maser in the galaxy
is of great interest with different angular resolution and different dense molecular gases at higher frequencies.
In this article, the results of high-resolution mapping of 22 GHz \ho maser and \hcn and \hco thermal molecular lines in NGC\,1068 using MERLIN and ALMA are presented in order to understand kinematics and conditions of gas in the circumnuclear region of the galaxy. 

Throughout this article, cosmological parameters of H$_{0}$ = 67.8 \kms Mpc$^{-1}$, $\Omega$$_{vacuum}$ = 0.692, and $\Omega$$_{matter}$ = 0.308 are adopted. The luminosity distance to NGC\,1068 is therefore 13.5 Mpc, and 1 arcsec corresponds to $\sim$65 pc in linear scale. 


\section{Observations and Data Reduction} 

\subsection{MERLIN Observations and Data Reduction}

Spectral-line observations at 22 GHz with the five telescopes of Multi-Element Radio-Linked Interferometer Network (MERLIN)\footnote{MERLIN is operated by the University of Manchester as a National Facility of the Science and Technology Facilities Council (STFC), and is available for use on a peer-review basis, with observing proposals dealt with by the Panel for Allocation of Telescope Time (PATT). 
The National Facility also hosts Jodrell Bank Observatory's VLBI participation, linking telescopes across Europe.}, including the Mark 2 telescope. 
The observations of NGC\,1068 were carried out over periods of 6-7 April in 2002 for $\sim$ 16 hours in total. 
Phase-tracking position for NGC\,1068 was 
RA 02$^{\rm h}$42$^{\rm m}$40.711$^{\rm s}$, Dec. -00$^{\circ}$00$\arcmin$47.810$\arcsec$ (J2000).
In our observations, a single IF band of 16 MHz width with opposite circular polarizations 
divided to 64 spectral channels, yielding a frequency resolution ($\Delta\nu$) of 250 kHz (or 3.4 \kmss) was centred at the one frequency setting and changed to other two frequency settings by shifting centre frequency during observations. 
Using this frequency configuration, we obtained three IF center frequencies, centred at \vlsr = 597, 899, and 1092 \kmss. The resultant velocity coverages are \vlsr = 490 -- 705,  790 -- 1005, and 985 -- 1195 \kmss. The velocity range of \vlsr = 705 -- 790 \kms~was not covered.
Phase-referencing observations were conducted at each frequency set, using a phase-referencing source 0237-027, an ICRF source (RA 02$^{\rm h}$39$^{\rm m}$45${\fs}$4723, Dec. -02$^{\circ}$34$\arcmin$40.913$\arcsec$ (J2000)). 
In earlier MERLIN observations, the position used for 0237-027 in J2000 is RA = 02$^{\rm h}$39$^{\rm m}$45$\fs$47226, Dec. =  -02$^{\circ}$34$'$40.$''$8984 ($\pm$ 10 mas) \citep{mux96}. 
Note that the differences of these positions are well within the MERLIN synthesized beam.
The observations were executed in a sequence of 6 min scans with cycling interval of 250 sec for NGC\,1068 and 70 sec for 0237-027.  

Amplitude and bandpass calibration were performed by observations of 3C\,286 and 3C\,273. Data analysis was conducted using the NRAO Astronomical Image Processing Software (AIPS) \citep{van96}.
After initial calibration and data editing, the phase reference source 0237-027 was self-calibrated, and the phase solutions obtained from the phase reference source were applied to all IFs containing line emission. 
In order to make a continuum map, spectral channels without line-emission were averaged.
Thus, imaging and CLEANing of the continuum emission was done separately.

The synthesized beam sizes (natural-weighting) used in the CLEAN deconvolution are 0.043$''$$\times$ 0.021$''$ in the position angle (P.A.) of 57$^{\circ}$, and the resultant rms noise level per a spectral channel ($\Delta\nu$ = 250 kHz) was about 5-6 \mb (A 0.02 arcsec synthesized beam corresponds to 1.3 pc at the galaxy.). 
The rms noise level of a line-emission-free map made by averaging line-emission-free channels was $\sim$ 2 \mb. 
In this article, the velocities are calculated with respect to the Local Standard of Rest (LSR). Imaging of the calibrate data was conducted with the AIPS.

\subsection{ALMA Data and Data Reduction}

For analysis of the thermally-excited molecular lines and the sub-millimeter continuum, we used data of ALMA band 6 (211--275 GHz) observations (Project: 2016.1.00052.S, PI: M.Imanishi) conducted at higher spatial resolution. 
The observations were made from 2016 October to 2017 September.
For these Cycle 4 observations, 40-44 12 m telescopes were used as a hybrid configuration with baselines ranging from 17 to 7552 m. 
This configuration results in a largest recoverable scale of $\sim$1.3" for extended emission and an angular resolution of  0.02".
In this analysis, two 1.875 GHz wide spectral windows were used to cover  \hco ($J$=3--2) and HCN ($J$=3--2) emission lines at 267.558 GHz and 265.886 GHz, respectively.  
More details of the observations, calibrations, and data analysis are described in \citep{ima18}.  
We analyzed \hco ($J$=4--3) emission line data at 356.734 GHz (Project: 2016.1.00232.S) obtained from the ALMA Science Archive.  
The observation was conducted on 2016 September 8.

Imaging of the calibrated data was carried out using the Common Astronomy Software Applications (CASA) \footnote{\url{https://casa.nrao.edu}}\citep{casa22}.
The continuum emission of the galaxy was subtracted from the calibrated spectral line visibility cubes using line-free channels prior to CLEAN deconvolution of line emission. The emission lines were separated out from the continuum emission. Imaging was performed using CASA with Briggs weighting (robust\,=\,0.5). 
The resultant synthesized beam size in the CLEAN is 0.03"--0.07".

\subsection{Effelsberg Observations}\label{ob3}
22 GHz \ho maser spectrum of the galaxies was obtained using Max-Planck Institute 100-m telescope at Effelsberg. The half-power beam width of the 100-m was $\sim$40" at 22 GHz.
The pointing calibration was made by observing 
nearby strong continuum sources every 1 to 2 hours, yielding 
a pointing accuracy of better than $\sim$ 5".
Observations of the maser were conducted using a 22 GHz-band receiver with a 300\,MHz bandwidth, 65536 spectral channels, and dual polarization from September 24 to 26 2022 at a spectral resolution of 4.6\,kHz or 0.063\,\kms. 
Smoothing 16-32 spectral points yields $\sim$ 1--2 \km velocity resolution at 22 GHz. Accuracy of flux density is estimated to be $\sim$30\%.

%

\begin{figure}
\centering
\includegraphics[width=1.15\columnwidth]{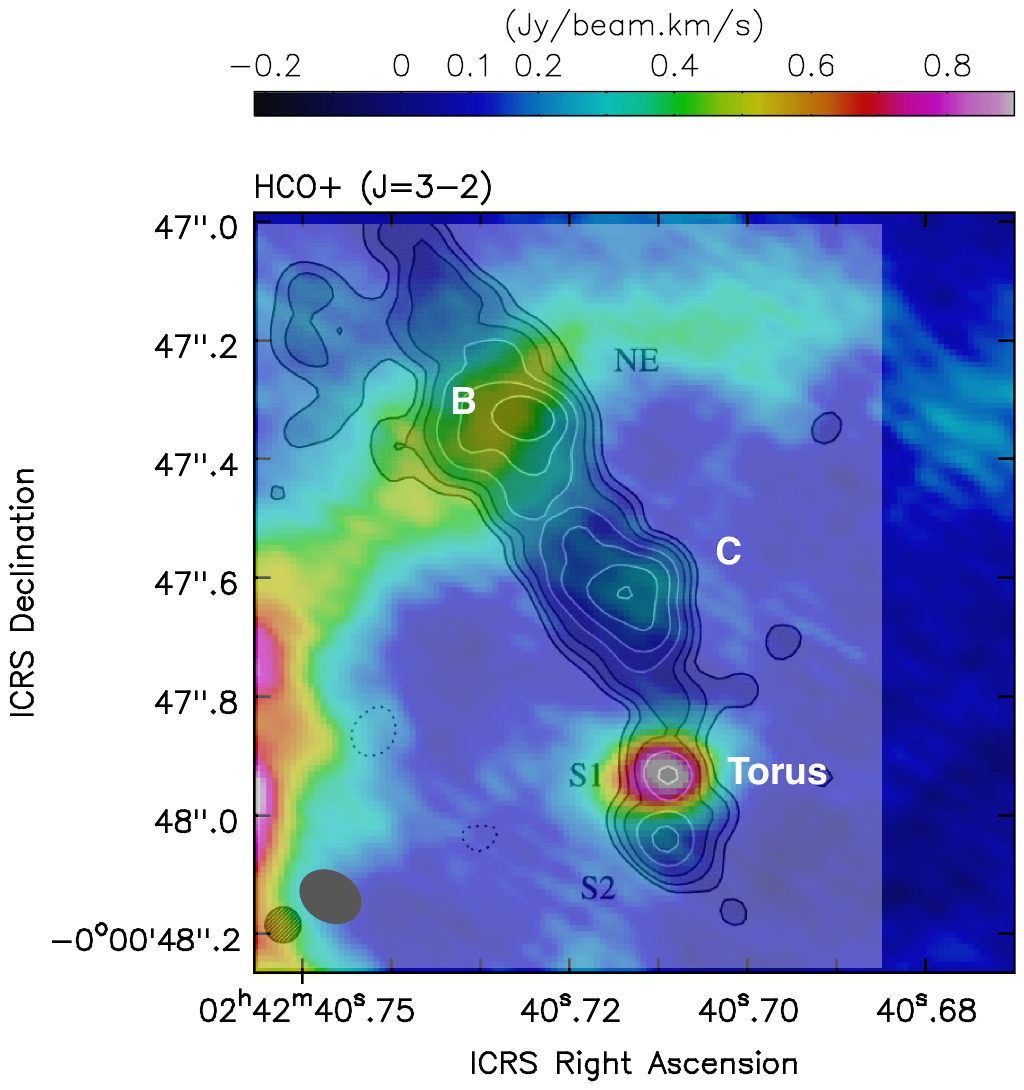} 
\vspace{-27mm}
\caption{MERLIN 5 GHz continuum contour map by \citet{gal04} overlaid the ALMA HCO (3--2) total intensity map appearing in Figure~\ref{f5}. The four major radio components NE,C,S1,and S2 and the molecular gas torus probed by HCN and \hco \citep{ima18} are indicated on the map. Location B, C, and S1 are \hco and \hcn emission peaks defined in Figure~\ref{f5}. The synthesized beams of MERLIN (smaller) and ALMA (larger) are plotted at the bottom left. }
\label{f1}
\end{figure}
\begin{figure*}
\centering
\includegraphics[width=1.5\columnwidth]{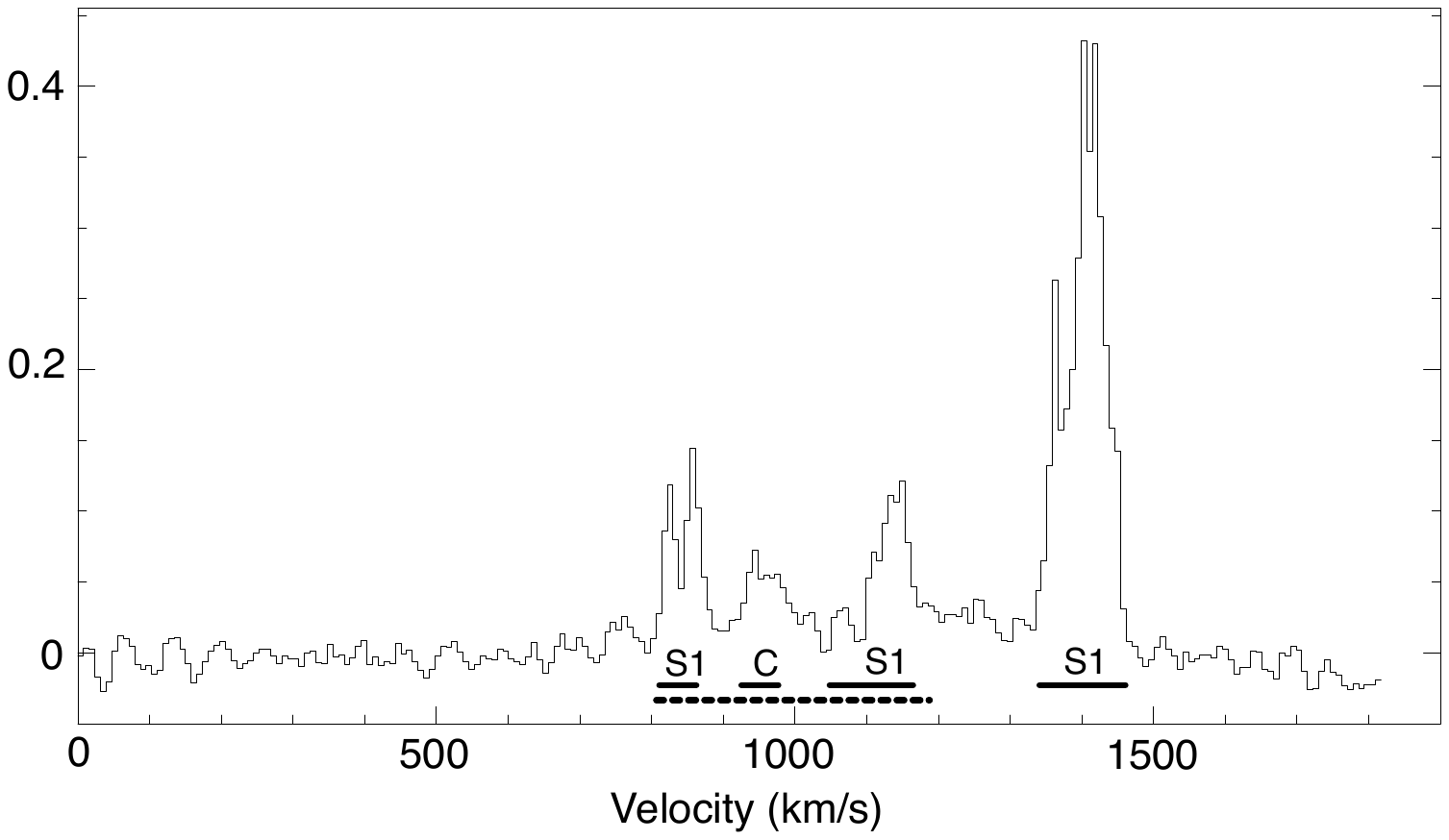}
\vspace{-5mm}
 \caption{22 GHz \ho maser spectrum towards the centre of NGC 1068, obtained at the Effelsberg 100-m telescope on 24-26 September 2022.  The systemic velocity of the galaxy is \vlsr=1123\,\kms (radio, LSR)is indicated by a vertical dotted line. 
 Short solid bars indicate the velocity ranges of the off-nuclear maser at component C and the nuclear masers at S1, respectively \citep{gal01}. A horizontal dotted lines indicate the velocity range observed by the MERLIN observation in this article. 
 Vertical axis indicates the total flux density in Jansky with the calibration error accurate to about 30 \%.}
  \label{f2}
\end{figure*}
%
\begin{figure}
\centering
\includegraphics[width=1.0\columnwidth]{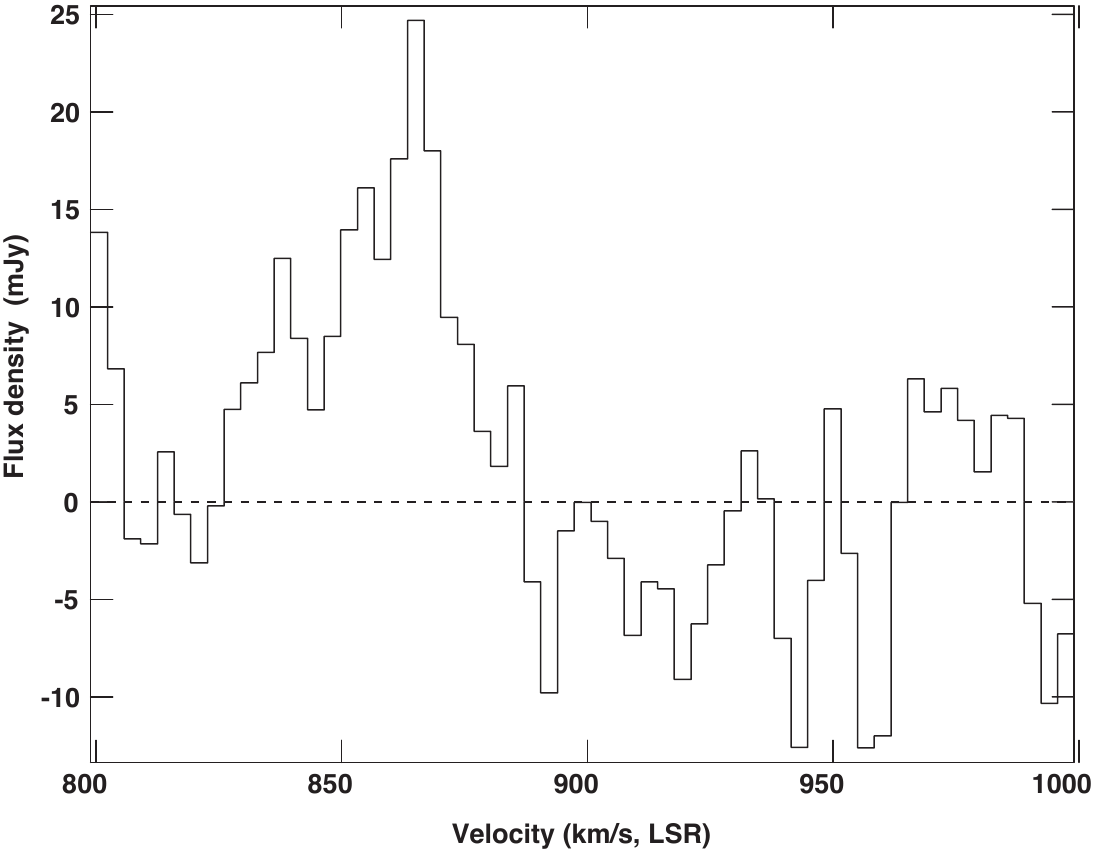}
 \includegraphics[width=1.0\columnwidth]{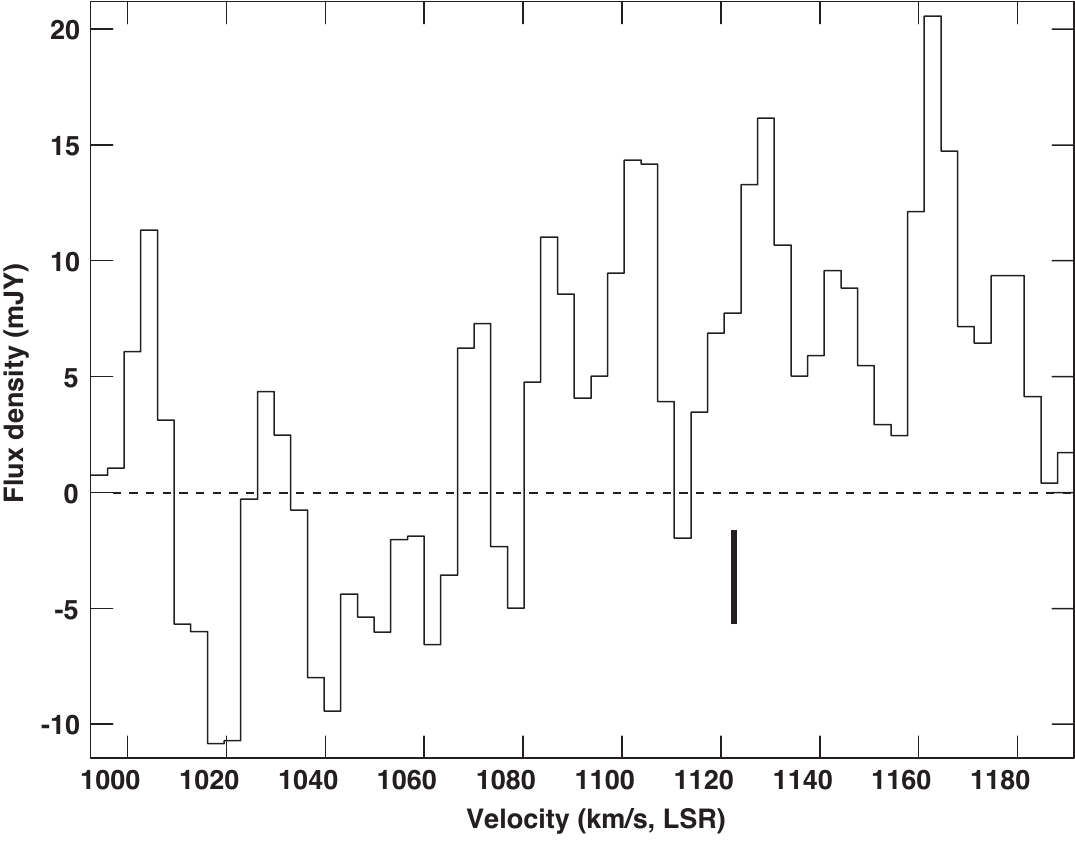}
\includegraphics[width=1.0\columnwidth]{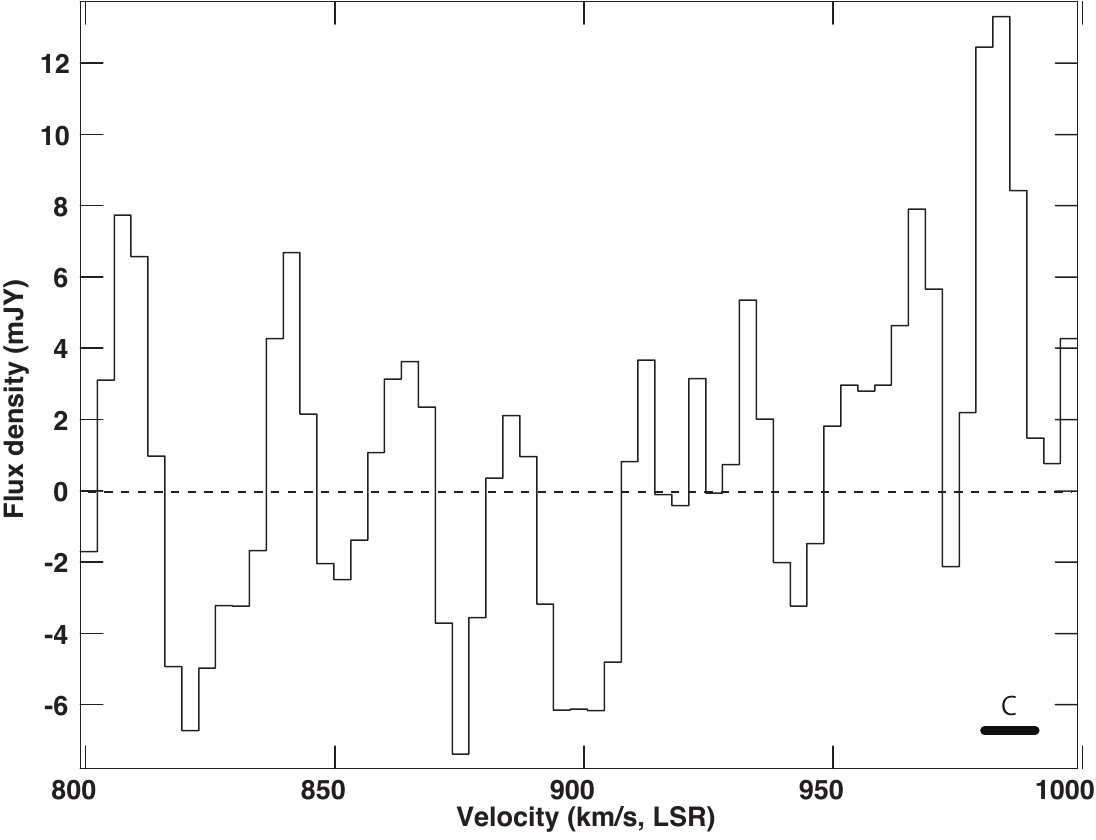}
\vspace{0mm}
\caption{Spectra of the 22 GHz \ho maser with Gaussian weighting, obtained with the MERLIN. 
These spectra show the nuclear masers towards S1 in the blue-shifted velocity range (top) and those in the systemic velocity range (middle). A vertical thick bar indicates the systemic velocity of the galaxy.  A spectrum of \ho maser towards the radio jet component C (bottom) shows a few minor peaks (denoted by a horizontal thick bar) centred at \vlsr= 980.4 \km .}
\label{f3}
\end{figure}
\begin{figure*}
\vspace{-30mm}
\centering
\includegraphics[width=15cm]{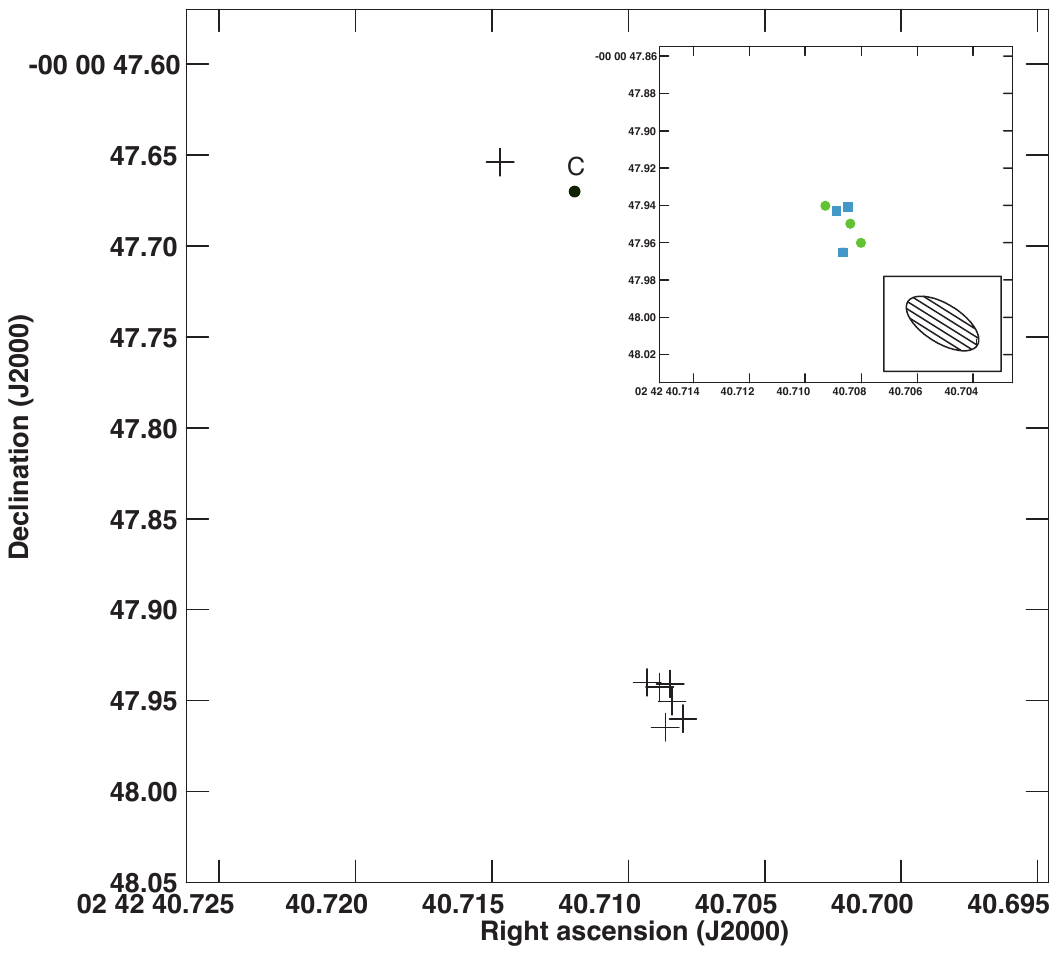}
\vspace{-45mm}
\caption{Centre positions of the nuclear masers and a tentative position of the off-nuclear maser, obtained by the MERLIN observation. A 22 GHz continuum peak in component C is marked by a black circle by assuming that positional uncertainties between the maser map by MERLIN and the VLA 22 GHz continuum map are negligible. The nuclear masers at the blue-shifted velocity (blue squares) and the systemic velocity (green circles) are shown also as an inset. The synthesized beam of 43.9 $\times$ 20.7 mas at P.A. =  58$\degr$ is plotted at the bottom right corner in the inset. }
\label{f4}
\end{figure*}
\section{RESULTS} 
\begin{figure*}
\centering
\includegraphics[width=19.8cm]{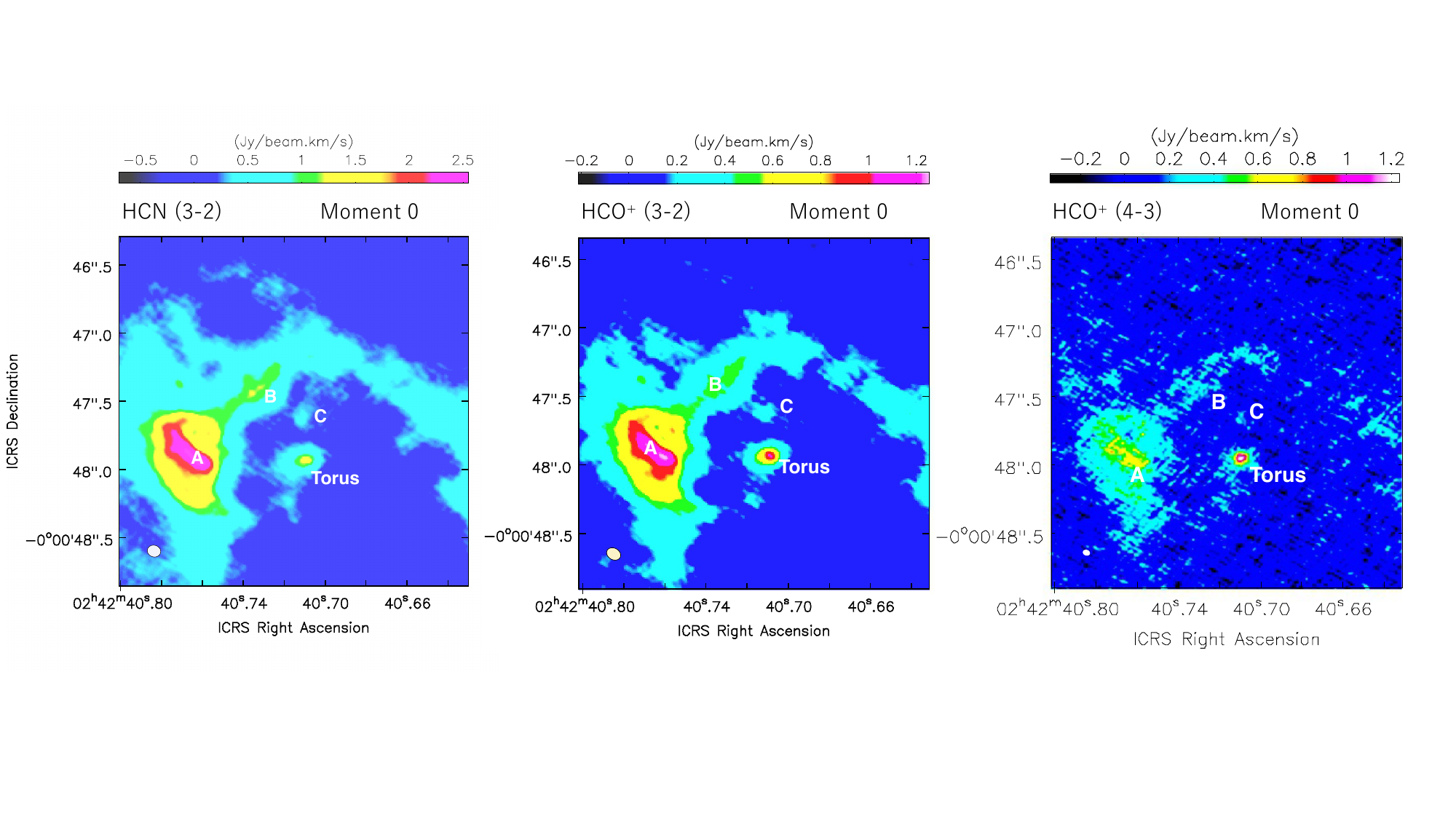}
\vspace{-25mm}
\caption {Color-scale total integrated intensity (zeroth moment) maps of \hcn and \hco line emission towards the circumnuclear region of NGC 1068, presented in \citet{ima18}. ($left$:) Zeroth moment map in HCN(3--2), ($middle$:) \hco(3--2) ($right$), and ($right$:) \hco(4--3). The components A, B, C (jet and off-nuclear maser), and the torus (AGN) are marked in each pannel. The synthesized beams of ALMA are plotted at bottom left. } 
\label{f5}
\end{figure*}

%
Figure~\ref{f1} shows the 5\,GHz continuum map of NGC\,1068 presented in \citet{gal04}, obtained at $\sim$ 0.1 $\arcsec$ $\times$ 0.05 $\arcsec$ angular resolution by MERLIN, overlaid on the total intensity map of the \hco (3--2) emission obtained by ALMA in Figure~\ref{f5}.
Figure~\ref{f2} shows 22 GHz \ho maser spectrum towards the center of the galaxy that was taken at the Effelsberg 100-m Telescope in September 2022.
Figure~\ref{f3} shows the maser spectra obtained by MERLIN towards the positions of components S1 and C in the velocity range of \vlsr = 800 $-$ 997 \km (the blue-shifted velocity range), \vlsr = 994 $-$ 1190 \km  (the systemic velocity range) towards the location of the nucleus S1, and  \vlsr = 800 $-$ 1000 \km towards the radio jet component C. 
The peak flux density of the maser is 30 \mb at \vlsr= 865 \km with a signal-to-noise ratio (SNR) of $\sim$5. 
The position of the maser at \vlsr = 1162 \km with a peak flux of 32 \mb~ near the systemic velocity is RA= 02$^{\rm h}$42$^{\rm m}$40.7093$^{\rm s}$, Dec.= -00$^{\circ}$00$\arcmin$47.9479$\arcsec$ (J2000). 
This coincides with the position of the AGN (RA = 02$^{\rm h}$42$^{\rm m}$40.71$^{\rm s}$, Dec. = -00$^{\circ}$00$'$47.94$\arcsec$ (J2000)) reported in \citet{ima18}.  
The \ho maser in the nuclear region of the galaxy was searched within the field of view of 0.768 $\times$ 0.768 arcsec$^2$, centred on the phase tracking centre of the galaxy and in the velocity range of \vlsr = $\sim$ 800 -- 1189 \kmss. 
We identified possible minor peaks straddling over a few spectral channels centred at \vlsr = 980.4 \kms. 
The centre velocity of the peak is similar to the known off-nuclear maser emission at V= 984.1$\pm$ 1.2 \kms \citep{gal01}.
The position of the off-nuclear maser is RA = 02$^{\rm h}$42$^{\rm m}$40.714$^{\rm s}$, Dec. = -00$^{\circ}$00$'$47.653$\arcsec$ (J2000), which is located near the radio continuum source C (RA = 02$^{\rm h}$42$^{\rm m}$40.715$^{\rm s}$, Dec. = -00$^{\circ}$00$'$47.636$\arcsec$ (J2000)) \citep{mux96}. 
Thus, the off-nuclear maser emission is tentatively found at  C,  despite the fact that the detection level is $\sim$2--3$\sigma$. 
The upper limit of the isotropic maser luminosity is estimated to be $\sim$ 0.1 $\lsun$.

The coordinates of the nuclear component at 5 GHz and the masers are listed in table~\ref{tab1}: A mean position of the detected masers in the blue-shifted and systemic velocity range is RA = 02$^{\rm h}$42$^{\rm m}$40.7087$^{\rm s}$, Dec. = -00$^{\circ}$00$\arcmin$47.949$\arcsec$ (J2000). 
Therefore, the positions of the masers in the blue-shifted and systemic velocity range are all confined within $\approx$ 0.02$\arcsec$  (20 mas) from the mean position, consistent with the results of earlier observations. 

Fig.~\ref{f4} displays the centre positions of the nuclear masers detected in an SNR of \ga 4 and a tentative position of the off-nuclear maser. 
In this figure, the nuclear maser positions are plotted also as an inset with uncertainties denoted by error bars.  
These nuclear maser features are peaked at \vlsr= 800, 861, and 865 \kms~in the blue-shifted velocity range and \vlsr = 1068, 1105, and 1189 \kms~in the systemic velocity range. 
All the detected masers remain unresolved at the angular resolution of $\sim$ 20 milliarcsecond (mas), or $\sim$ 1.3 pc.  
Considering the position error of the calibrator source ($\sim$5 mas), half the synthesized beam divided by SNR ($\sim$4 mas), the baseline errors  ($\sim$10 mas), and an error caused by an offset between the target and phase-reference source, positional accuracy is estimated to be $\sim$15 mas.
In our observations, no 22 GHz continuum has been detected to a 3 $\sigma$ rms noise of $\sim$ 6 \mb, making it difficult to compare precisely the locations of each maser spot and continuum components in the nuclear region of the galaxy. 

\begin{figure}
\centering
\includegraphics[width=0.9\columnwidth]{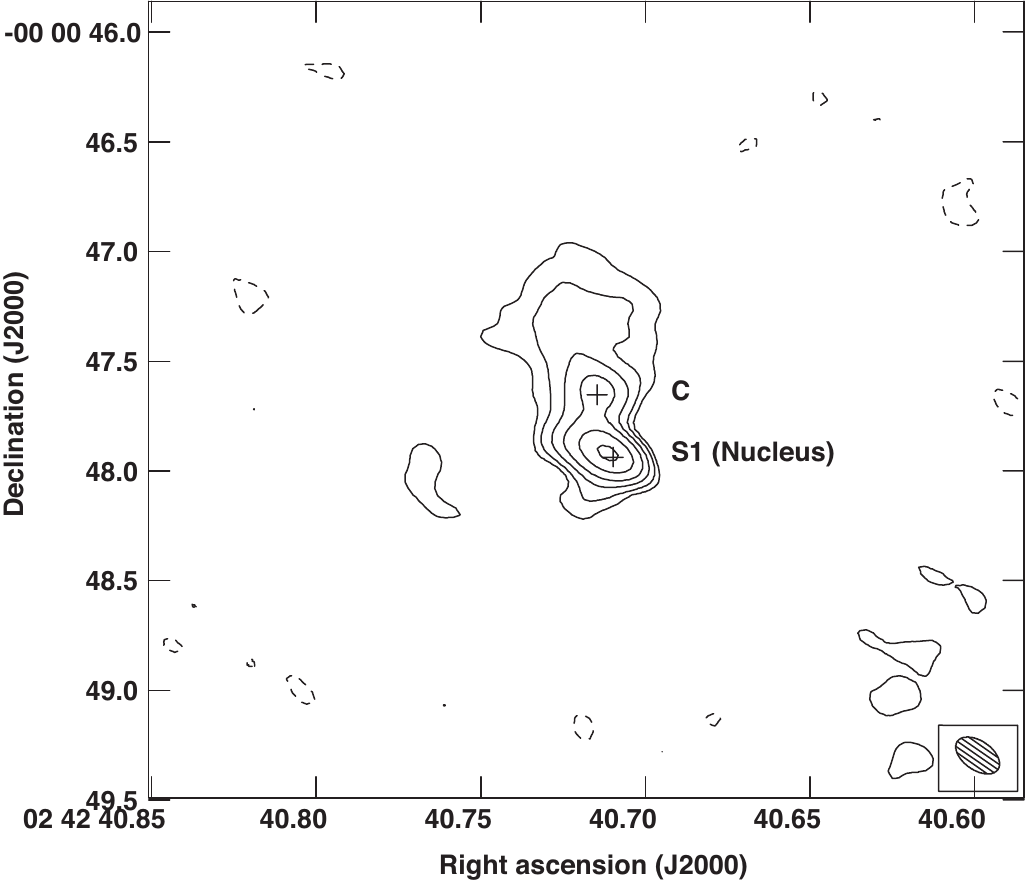}
\caption {Sub-millimetre (266 GHz) continuum map of a nuclear region obtained by the ALMA \citep{ima18}. 
Contours are -2,3,5.1,9,15,46$\times$1 $\sigma$ value of 0.18 \mb. 
The synthesized beam ($\sim$0.2$\arcsec$$\times$ 0.13$\arcsec$) is at the right bottom corner. 
Nuclear component of S1 at 5GHz and a tentative position of the 980.4 \kmss maser (Fig.~\ref{f2}) near the component C are marked by crosses.}
\label{f7}
\end{figure}

Fig.~\ref{f5} displays the velocity-integrated line intensity (zeroth moment) maps for \hco (3--2), \hco (4--3), and \hcn(3--2) emission and sub-millimetre (266 GHz) continuum emission is presented in Fig.~\ref{f7}, obtained from the identical spectral-line cube in Fig.~\ref{f5}.

\begin{figure*}
\centering
%
\includegraphics[width=18cm]{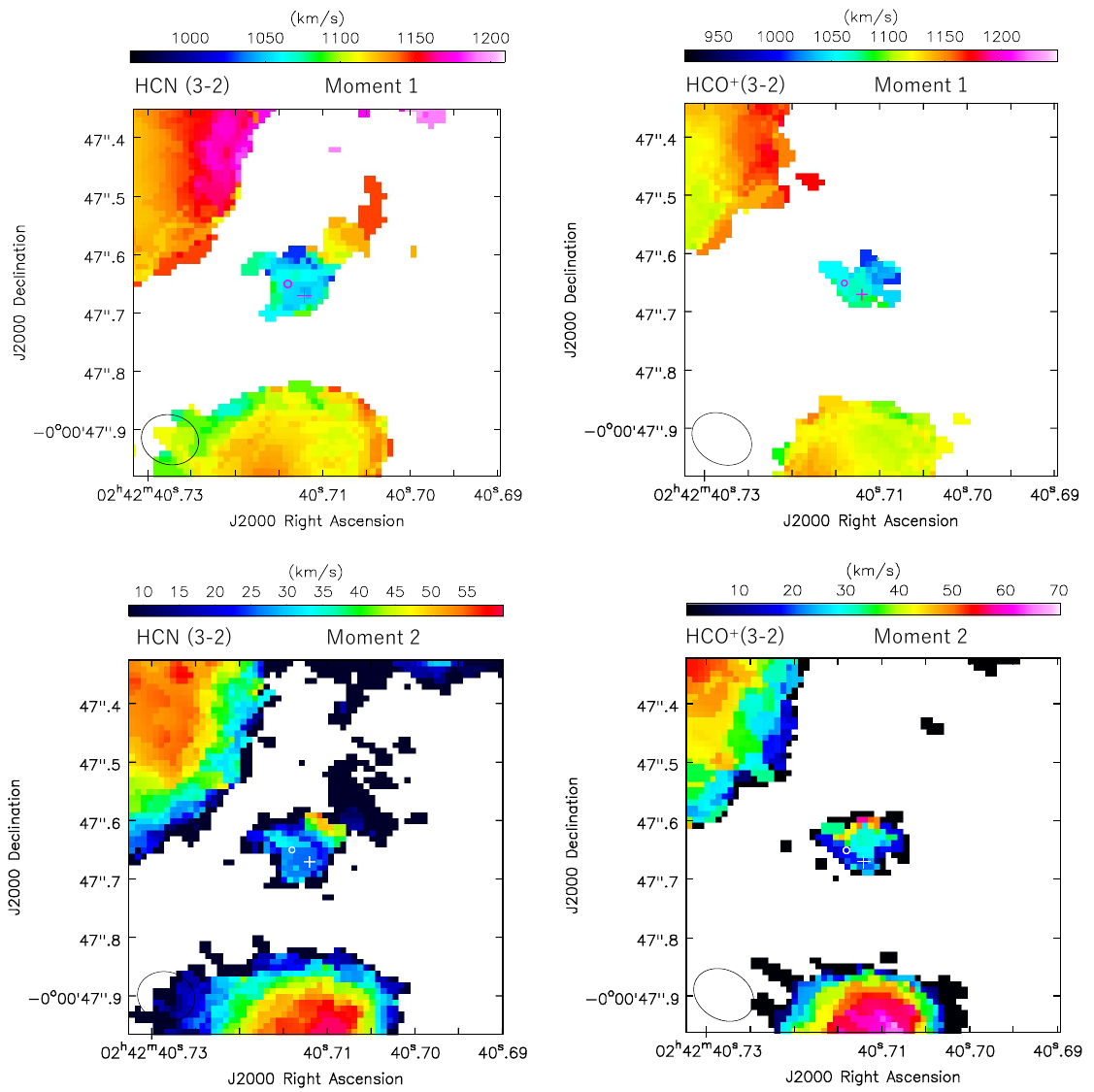}
\caption{Moment maps in HCN (3--2) and \hco (3--2) towards the nuclear region of NGC 1068, obtained from the ALMA Cycle 4 observations.
Mean velocity (first moment) maps ($upper$), and velocity dispersion (second moment) maps ($lower$) in HCN (3--2)  and \hco (3--2) are presented. A 22 GHz radio continuum peak position of component C is denoted by a cross and the blue-shifted maser spot (\vlsr=980\kms) detected by MERLIN is marked by a circle in each panel.).
The synthesized beam sizes ($\sim$0.04") are shown in the bottom left of each panel. In these maps, the optical convention for Doppler velocities are adopted. The difference between the optical and radio velocity conventions is $\approx$ 5 \kms. The coordinates in each panel are based on the J2000 system.}
 \label{f6}
\end{figure*}

Fig.~\ref{f6} displays mean velocity (first moment) and velocity dispersion (second moment) maps in \hco (3--2) and \hcn(3--2) molecular lines.

Table~\ref{tab2} presents line intensity ratios of \hco (3--2) to \hco (4--3) and \hcn (3--2) to \hco (3--2) at locations A, B, C and, S1 appearing both in Figure~\ref{f1} and Figure~\ref{f4}.
%
%
\begin{table}
	\begin{centering}
\caption{Positions of radio continuum and water maser emissions in NGC\,1068. }
	\label{tab1}
\begin{tabular}{lccc} 
		\hline
Component & RA(02$^{\rm h}$42$^{\rm m}$)& Dec(-00$^{\circ}$00$\arcmin$)& $\sigma_{p}$$^{a}$\\
&(J2000)&(J2000) \\
		\hline
5GHz$^{b}$ \\
~~~MERLIN\\
~~~~S1 (Nucleus) &40.7098$^{\rm s}$ & 47.938$^{\arcsec}$& 0.02$^{\arcsec}$\\
~~~~C (Jet) &40.715$^{\rm s}$ & 47.636$^{\arcsec}$& 0.02$^{\arcsec}$\\
~~~VLBA\\
~~~~S1 (Nucleus) &40.70907$^{\rm s}$ &47.9445$^{\arcsec}$& 0.0004$^{\arcsec}$\\
~~~~C (Jet) &40.71298$^{\rm s}$ & 47.6577$^{\arcsec}$& 0.0005$^{\arcsec}$\\
\hline
22 GHz$^{c}$ \\
S1 (Nucleus) &40.709$^{\rm s}$ & 47.94$^{\arcsec}$& --\\
C (Jet)&40.712$^{\rm s}$ & 47.67$^{\arcsec}$& --\\
		 \hline
Nuclear Water Masers &&& $\sim$0.01$^{\arcsec}$\\
\vlsr=865\kms   & 40.709& 47.94&   \\
~~~~~~~=861\kms   & 40.708 & 47.96&   \\
~~~~~~~=861\kms   & 40.708 & 47.96&   \\
~~~~~~~=865\kms   & 40.708 & 47.94&   \\
~~~~~~~=1105\kms  &  40.708& 47.95&  \\
~~~~~~~=1162\kms  &  40.709& 47.94& \\ \\
Mean position &  40.708& 47.94& \\
    \hline 
Off-nuclear Maser$^{d}$&  & & \\
\vlsr=980\kms   & 40.714 & 47.65&   \\
		\hline 
	\end{tabular}
 \end{centering}
  {$^{a}$ Position errors. The errors in Dec are larger due to the low declination of the galaxy.}\\
 {$^{b}$5-GHz nuclear continuum positions and errors are from \citet{mux96} (MERLIN) and \citet{gal04}  (VLBA).}\\
 {$^{c}$Measured at 22 GHz in 2015 by VLA \citep{mut24}.}\\
 {$^{d}$ Tentative detection}

\end{table}
\begin{table}
	\begin{centering}
\caption{\hco and \hcn Line integrated intensity (in Jy \kms) ratios at Components A, B, C, and S1 are listed. }
	\label{tab2}
\begin{tabular}{lcc} 
		\hline
Component & \hco(3--2)/(4--3)$^{\ast}$ &  \hcn(3--2)/\hco(3--2) \\
		 \hline
		 A   &1.25  & 2.45  \\
		 B  & 1.75  &  1.66\\
		 C  &  0.87 &  1.35 \\
		 S1(Torus)&0.83  & 1.33 \\
			\hline 
	\end{tabular}
\end{centering}
{$^{\ast}$The \hco(4--3) emission data from Figure~\ref{f5}}
\end{table}
%


\begin{figure}
\centering
\includegraphics[width=1.10\columnwidth]{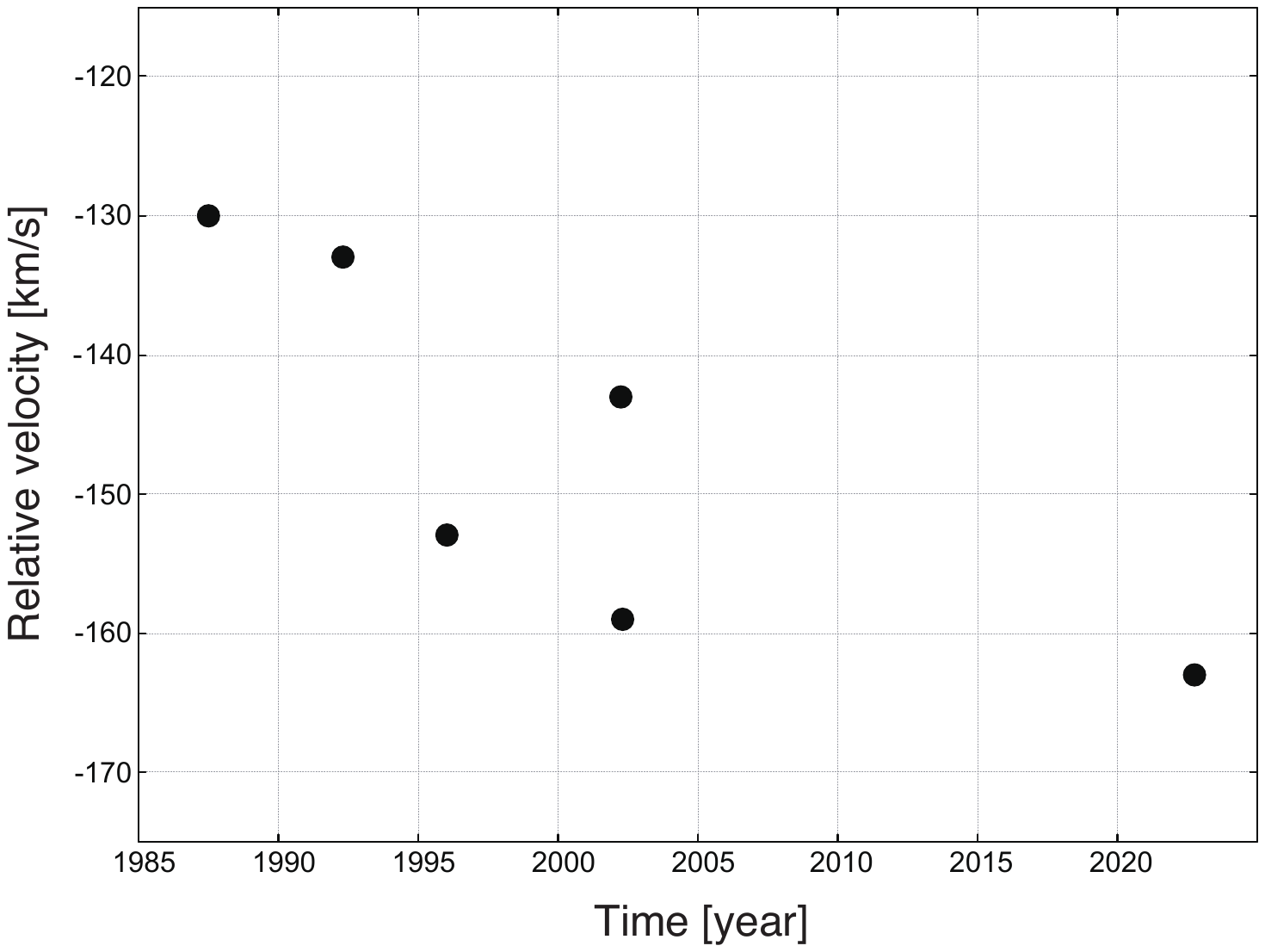}
\caption{Time-series of the centre velocities of the off-nuclear \ho maser obtained from 1987 to 2022. The vertical axis denotes centre velocities of the maser that are relative to the systemic velocity of NGC\,1068, which are estimated from the spectra in the following references: \citet{gal96}, \citet{nak95}, \citet{gal01}, \citet{bra03}. and this article. 
The horizontal axis is time scaled in year.}
\label{f8}
\end{figure}
%
\section{DISCUSSION} \label{sec:discussion}
\subsection{Off-nuclear maser}

In our MERLIN observations, the off-nuclear \ho maser at component C was not distinctly detected. 
It is likely that the flux density of the maser was weaker when observed or it was resolved out by our MERLIN synthesized beam that is by a factor of 5 smaller than that for earlier detection with the Very Large Array (VLA) \citep{gal96}. 
However, the latter is less likely since the masers lying in the blue-shifted and systemic velocity ranges were detected in our MERLIN maps. 
Moreover, according to the recent research using a global VLBI network including the National Radio Astronomy Observatory (NRAO) VLBA, the phased-VLA, and the Effelsberg 100-m, which was observed on 2000 February 23-24,  the \ho masers at C were detected at an even smaller 1.3 $\times$ 1.0 mas angular resolution (see Figure~\ref{f9})\citep{mor22}.  
The off-nuclear masers at C were previously detected in velocities of \vlsr $\simeq$ 950--1050 \kms by the VLA in 1987, and the flux density of the maser was $\sim$ 50 mJy \citep{gal96}.  
If the strength of the maser was as strong as that observed by the VLA, the maser should have been clearly detected in our observations.  
Nevertheless, a minor 2.5$\sigma$ peak at \vlsr = 980.4 \kms seen with MERLIN near component C in the Figures~\ref{f3} and \ref{f4} would represent the location of the off-nuclear maser. The observed intensity variability of the off-nuclear maser \citep{gal01} may indeed account for the reduced flux density of the maser at component C below $\sim$20 mJy at the time of our MERLIN observations.
%



\subsection{Nuclear \ho masers at the continuum nucleus (S1)}

One important result in our observations is that the positions of the nuclear masers in NGC\,1068 obtained at the angular resolution of $\sim$ 20 mas have been estimated with respect to the radio nuclear component S1 in Figures~\ref{f1} and \ref{f5} from earlier 5 GHz MERLIN data and has not been well established by VLBI observation.
Earlier VLBI observations at 22 GHz \citep{linc96} did not detect the nuclear continuum components, which makes astrometric registration difficult for the masers and the radio nucleus. 
An attempt has been made to register the relative positions of all maser emissions extending to $\pm$300 \kms below systemic and the radio continuum components using VLBI data at 5 GHz \citep{gal04}.
The positional uncertainties resulting from the not precisely known position of the radio nucleus S1 are comparable with those of the masers.
The positions of the blueshifted masers presented in \cite{linc96} were not obtained with phase-referencing observations and their positions were estimated relative to the brightest maser spot in the galaxy \cite{gal01}.
In our observations, the 22 GHz continuum was not detected and the positions of the nuclear masers have been determined with respect to the phase-referencing source used in both our observations and earlier MERLIN observations \citep{mux96}. 
The coordinates of the nuclear component at 5 and 22 GHz and the masers are listed in Table~\ref{tab1}.
The positions of the unresolved maser sources and the continuum nucleus at 5 GHz are co-located within the uncertainties. 
It should be noted that the maser positions are offset south-west or south with respect to the 5 GHz or 22 GHz continuum nucleus 
and some systematic position shifts between the maser spots and the 5 GHz S1 continuum source could not be removed in our data analysis.
Nevertheless, we conclude that the detected blue-shifted and systemic masers arise near from the central engine of the galaxy, which is consistent with the results of earlier VLBI measurements.





\subsection{Velocity fields in \hcn and \hco at C}

The velocity gradients in both the \hcn (3--2) and \hco (3--2) emission at the radio component C have been resolved at the $\sim$0.03$\arcsec$ spatial resolution of ALMA in Figure~$\ref{f6}$. 
As is evident in the first moment maps of Figure~\ref{f6}~(upper), the velocity gradients in the \hcn and \hco lines are spanning \vlsr = $\sim$1050 to 1150\,\kms and \vlsr = $\sim$1000 to 1100\,\kms, respectively. 
However, the directions of the gradients are seen to be opposite:  the \hcn gradient appears increasing towards north-west with red-shifted by $\sim$100 \kms, while the \hco gradient decreases westward with blue-shifted by $\sim$100 \kms. 
The latter seems to trace outflowing materials in the local circumnuclear medium that would be with the velocity distribution of the \ho maser emission imaged at C with milliarcsec VLBI observations as discussed below. 
However, there is no straightforward interpretation for this discrepancy except for the possibility that the \hcn samples a much larger emission component. 

Figure~\ref{f6}~(lower) shows the second moment maps for the two molecular emissions. 
In the figures, we find a high velocity dispersion region with a velocity dispersion of $\sigma_{v}$  = $\sim$30--50~\km in \hcn and $\sigma_{v}$ = $\sim$30--60~\km in \hco. The \hco dispersion also shows a small gradient from nearly south to north along the assumed trajectory of the jet. 
The high-dispersion region in \hco appears offset about 0.05" or 3.3 pc north of the center of C, and the high-dispersion regions in both \hcn and \hco are located at the edges of the line emission regions. 
This may indicate that the jet-ISM interaction creates a cavity that forces gases to accumulate at the edges of the gaseous regions.

\subsection{Systematic velocity shift of the off-nuclear maser}

Figure~\ref{f8} shows time-series of the centre velocity of the off-nuclear maser at six epochs spanning from 1987 to 2022 covering a period of 35 years. 
To determine the centre velocity, the strongest single-peak line emission was used at each epoch, and these emission line profiles did not significantly change over the period.
The values of the velocities at each epoch are taken from \ho maser spectra in the literature listed in Fig.~\ref{f8} caption. 
It should be noted that the flux density of the off-nuclear masers are very variable and that the maser features at C at around \vlsr=$\sim$980 \kms were not always distinctly seen in the single-dish spectra \citep[e.g.,][]{bra03}.
Despite the fact that the measured velocity-shift shows such ambiguities, the plot does suggest a systematic velocity shift of about 0.85 km s$^{-1}$ yr$^{-1}$ in velocity in time towards the blue. 

\begin{figure*}
\centering
\includegraphics[trim={11.5cm 0cm 1cm 2.5cm},clip,width=21cm]{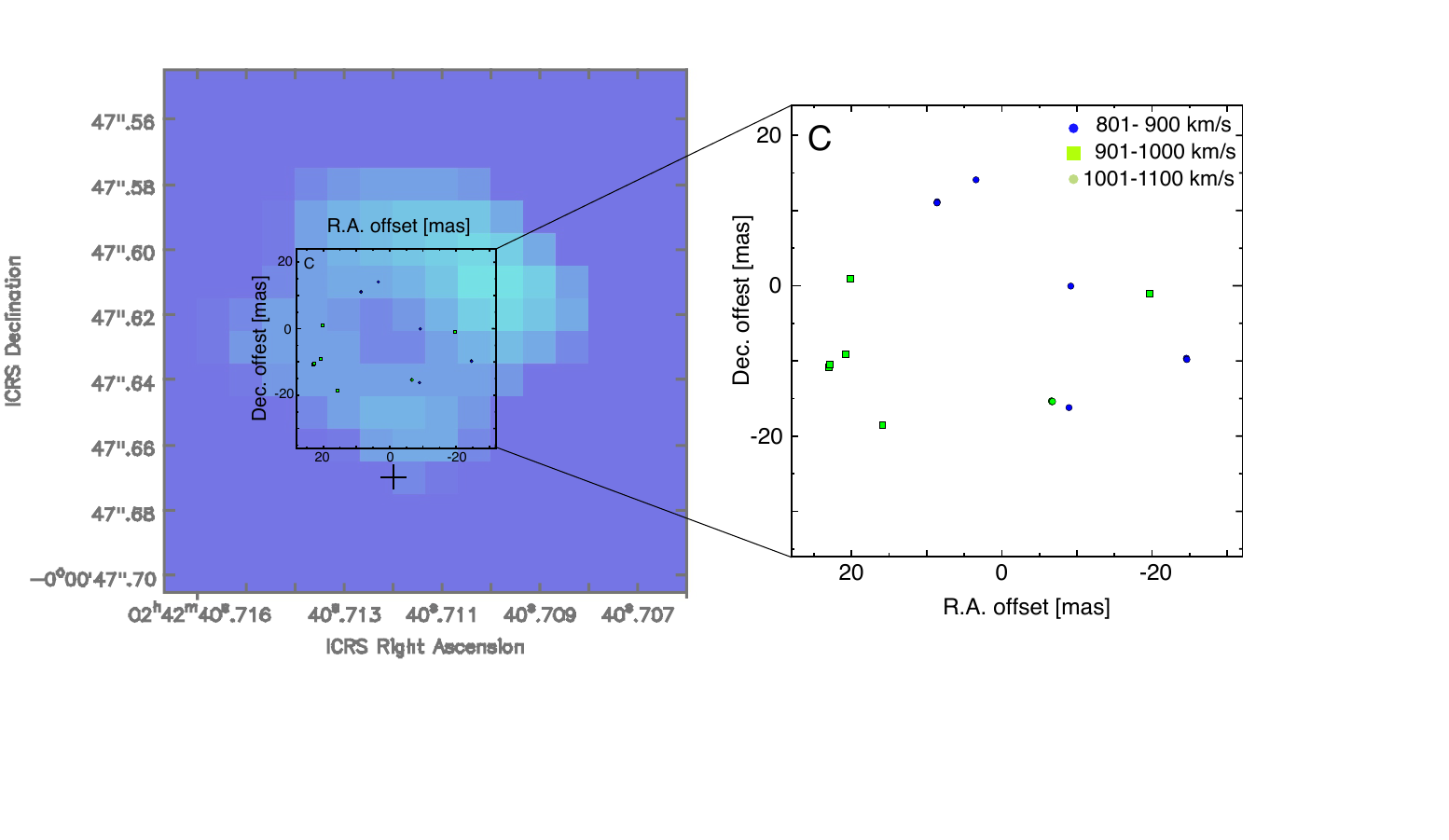}
\vspace{-20mm}
\caption{($Left$:)The off-nuclear maser spots superposed on the \hco(3--2) total intensity map (Figure~\ref{f5}) (Sudou, H., private communication). The picture assumes that the centre of the maser distribution corresponds approximately with the centre of the \hco (3--2) map having a hole or a cavity in the middle. The 22 GHz radio peak position measured in component C measured by VLA is marked by a cross (Mutie, Beswick et al. in preparation).  ($Right$:) Zoom-up map of the off-nuclear maser spots obtained by VLBI \citep{mor22}. The  \vlsr=980 \kms~ off-nuclear maser feature in Fig.\ref{f4} could be found in a group of the features lying at \vlsr=901-1000 \kms (green squares).}
\label{f9}
\end{figure*}

Figure~\ref{f9} shows the distribution of the \ho maser spots at component C obtained by VLBI that is superposed on the \hco (3-2) emission in Figure~\ref{f5}, assuming that the centre of the maser distribution largely corresponds to that of the \hco(3--2) ring-like structure.
The maser image has been obtained with a high-sensitivity array (HSA) consisting of the VLBA, the phased-VLA, and the Effelsberg 100-m telescope on 2000 February 23-24 \citep{mor22}. 
This figure places the maser component structure around the cavity found in \hco (3-2), although there are no absolute coordinates registered for the maser distribution map presented as compared with the \hco map in \citet{mor22}. 
Note that the 22 GHz continuum peak position of component C (marked by a cross in Figure~\ref{f6}) lies nearly south of the HCN and \hco emission, which implies that the continuum peak lies at the shock front between the jet and the intervening molecular gas. 
It is interesting that the overall distribution of the masers at component C shows a ring-like structure with a hole or cavity at its centre, which may indicate that the presence of a strong shock in the jet-ISM interaction causes a ring-like \hco emission structure. 
The blue-shifted maser features spanning \vlsr=900--1100 \kms depicted in Figure~\ref{f9} provide evidence that the shock front between the jet and the intervening molecular gas is steadily moving towards the observer.  \\

Very recently,  the kinematics of the maser at C has been revealed by the HSA observations, in which the off-nuclear masers appear to trace an expanding ring structure \citep{gal23} similar to the ring-like distribution of the \hco emission in our data (Fig.~\ref{f9}). Assuming that the radius of the ring in Fig.~\ref{f9} is 20 mas (1.3 pc) and the expansion speed is on the order of 100 \kms, based on the outflowing velocity shown in Fig.~\ref{f8}, or the outflow velocities  in the shocked interstellar medium in AGN \citep{pou13}, the kinetic age of the ring may be estimated at $\sim$12000 years. This value is similar to the 6000 years obtained from the Expanding Ring Model presented in \citet{gal23}.
Moreover, the positions of the masers are displaced $\sim$5 mas south of the 5 GHz continuum peak at C in the new HSA data, and the blue-shifted maser spot (Fig.~\ref{f4}) is displaced $\sim$2 mas east of the 22 GHz continuum peak. Both displacements could indicate that the jet advances in the molecular medium and the maser inversion occurs at the jet shock front.

\begin{table}
	\centering
\caption{The relative velocities of the off-nuclear masers as a function of time are presented. 
The relative velocities are estimated from the difference between the systemic velocity of the galaxy and the centre velocity of the maser emission at each epoch. 
The value at the 1996.0 epoch is from the integrated spectra spanning 1995-1998 as stated in \citet{gal01}. 
Data points appearing in Figure~\ref{f8} are from this table.}
	\label{tab3}
	\begin{tabular}{lcc} 
	\hline
	Time & Relative & References \\
         & Velocity &   \\
    (Year) & (\kms) & \\
	\hline
	1987.50   & -130   & \citet{gal96}  \\
	1992.33   & -133   & \citet{nak95}  \\
    1996.0   &  -152.9 & \citet{gal01}  \\
    2002.26   &   -143 &  This article (Fig.\ref{f3}) \\
    2002.32   &  -159  & \citet{bra03}  \\
    2022.76   &   -163 &  This article (Fig.\ref{f2})  \\
	\hline 
	\end{tabular}
 \end{table}
 
\subsection{Conditions of the molecular medium at C}
\begin{figure*}
\centering
\includegraphics[trim={0cm 0cm 0cm 0cm},clip,width=8.5cm]{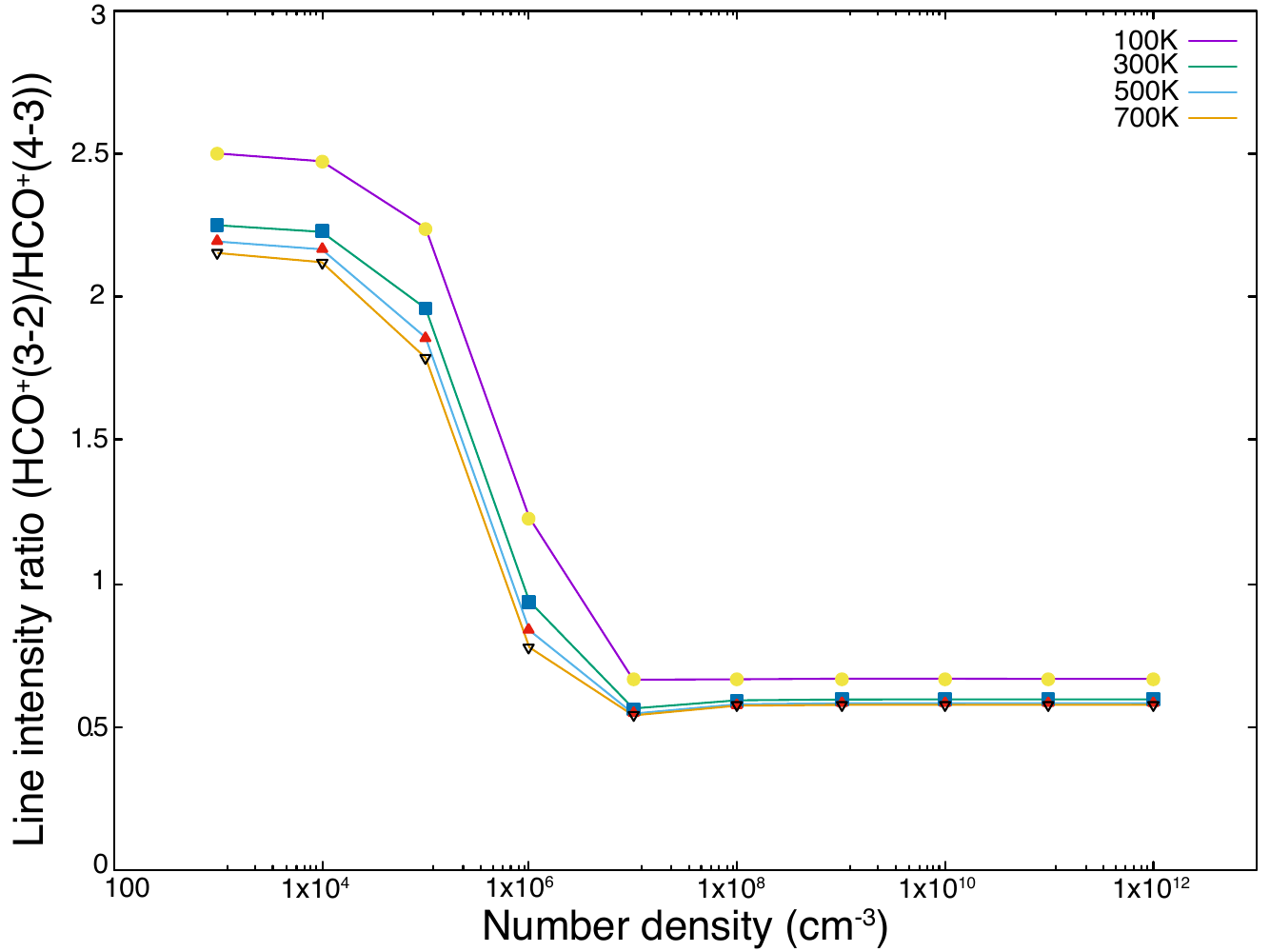}
\includegraphics[trim={0cm 0cm 0cm 0cm},clip,width=8.5cm]{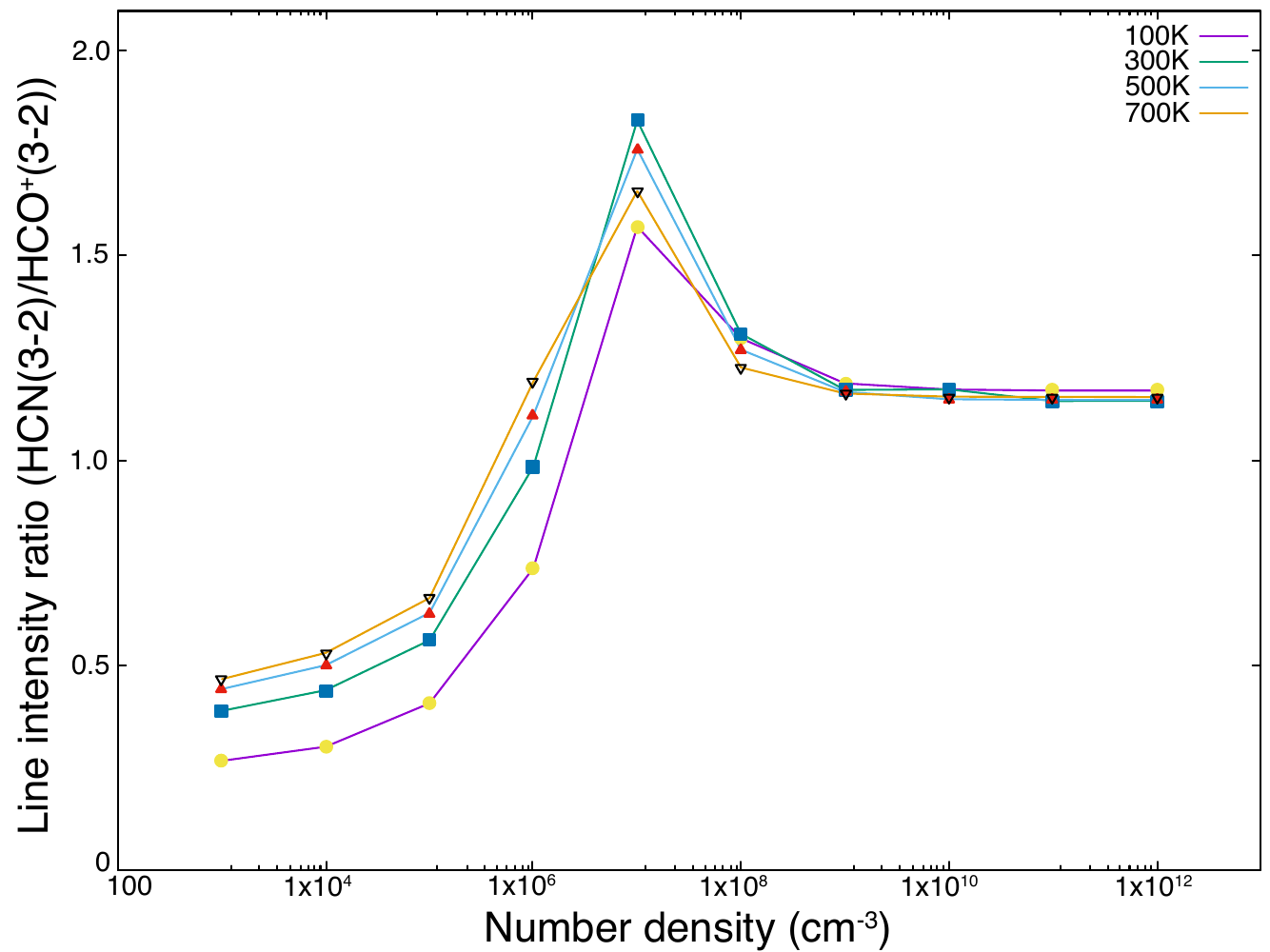}
\caption{($Left$): The \hco(3-2)/\hco(4-3) integrated flux density ratio as a function of the hydrogen number density (\nhd in cm$^{-3}$) for three different kinetic temperatures: T$_{\rm kin}$=100, 300, 500, and 700 K, assuming the cosmic microwave background temperature(T$_{\rm bg}$)=2.73 K, hydrogen column density(\nhh) = 10$^{13}$ cm$^{-2}$, and a line velocity-width($\Delta$V)=100 \kms. ($Right$): The \hcn(3--2)/\hco(3--2) integrated flux density ratio as a function of \nhd for T$_{\rm kin}$=100, 300, 500, and 700 K, assuming the \hcn to \hco molecular abundance ratio of 2. For other parameters, the same values as those in the left figure are adopted.
The results in these plots are calculated using the RADEX program \citep{tak07}.}
\label{f10}
\end{figure*}

The observed line intensity ratios of the \hco(3--2) and \hco(4--3) and \hcn(3--2) and \hco(3--2) at component C show values that are very different from those at components A and B in Figure~\ref{f1}. 
A non-LTE radiative transfer calculation based on the RADEX program \citep{tak07} indicates that the observed integrated flux ratio of \hco(3--2) to \hco(4--3) (0.87) in Table~\ref{tab2} can be explained by a hydrogen number density (\nhd) of $\approx$10$^{6}$ cm$^{-3}$ and a kinetic temperature of T$_k$=300--500 K or even warmer (Fig.~\ref{f10}), assuming a hydrogen column density of 10$^{13}$--10$^{15}$ cm$^{-2}$ and an \hco line-width of 100 \kms. 
These resultant values are consistent with the conditions (T$_{k}$=400-1500 K, \nhd \ga 10$^{6}$ cm$^{-3}$) where water maser excitation occurs \citep[e.g.,][]{gray16,gray22}. 

The enhanced line intensity ratio of the \hcn(3--2) and \hco(3--2) lines at C (1.35 in Table~\ref{tab2}) could be explained under conditions of T$_{k}$ $\geq$ 300 K, \nhd$\ga$10$^{8}$ cm$^{-3}$ if we take the molecular abundance ratio of \hcn to \hco to be about 2 (Fig.~\ref{f10}). 
Enhancement of the HCN to \hco abundance ratio by a factor of three or even larger than 10 has indeed been found in nuclear AGN environments, although this may not be representative for the off-nuclear region C in NGC\,1068 \citep[e.g.][]{izu16}. 
On the other hand, studies of the molecular abundance ratio in Ultra-luminous infrared galaxies (ULIRGs) show that the HCN to \hco flux ratio also increases with changes of the HCN/\hco abundance ratio as a function of kinetic temperature and hydrogen density \citep{baa10,ima23}.  In previous research, an elevated ratio of HCN to \hco has been observed not only in the nuclear regions of AGNs and starburst galaxies but also in regions where shocks occur, such as in molecular outflow from star-forming regions \citep{bac97}. Mechanical heating by shocks in nuclear molecular outflows in ULIRGs could also cause the enhanced intensity ratio \citep{aal12,izu13,ima19}.
%
\subsection{Non-detection of the sub-millimetre maser and continuum emission}
Figure~\ref{f7} shows that the submillimetre continuum emission at 266 GHz towards the nuclear region of the galaxy, in which the Component C is peaked north of the nuclear continuum peak S1. The Spectral Energy Distribution of the galaxy suggests that the continuum at 266 GHz is dominated by dust emission (using the photometric data points of NGC\,1068 from NED) and the continuum component traces the dust emission from foreground material and heated by the obscured active nucleus.

Water maser emission at both 183 and 321 GHz would serve as a further diagnostic of the local environments but these emissions have not been detected towards the nuclear region and component C in NGC\,1068 \citep{hagi16,pes16}.
The energy level of \ho in the 183 GHz transition ($E_u/k$=205 K) is lower than those of the 22 GHz (644 K) and 321 GHz (1862 K) transitions.
Assuming that emissions in these higher transitions also result from maser amplification of the background dust emission by foreground material, recent ALMA observations at $\sim$0.6" angular resolution have not detected the 183 GHz maser transition at a 5$\sigma$ upper limit of 37 mJy above a detected background continuum flux of 128 mJy arcsec$^{-2}$ \citep{pes23}. 
Considering the small ratio of the upper limit of the maser flux and the continuum, low-gain amplification could  result in observable maser emission.
%
%





\section{Summary}

MERLIN observation of the 22 GHz \ho maser emission in NGC\,1068 at $\sim$ 20 mas angular resolution has been conducted to resolve the maser and to understand the kinematics of the dense molecular gas at the shock front between the radio jet and intervening gas in the galaxy.
The nuclear masers at blue-shifted and systemic velocities were detected at the location of the 22 GHz continuum nucleus of the galaxy within uncertainties.  
Only a minor maser peak was seen at \vlsr = 980.4 \kms at the known off-nuclear maser position at component C, which would be consistent with the earlier results obtained by the VLA in 1983 and 1987. 
This maser may have faded during our observing period in 2002. 

Molecular emission lines of \hco at 257\,GHz and \hcn at 255\,GHz show an enhanced dispersion at C and weak velocity gradients in the roughly southeast to northwest direction.
However, the direction of the velocity gradients of \hco and \hcn is almost opposite, which may suggests that each of the emission lines probe different molecular structures.
The \hco velocity gradient is largely consistent with the weak velocity gradient traced by the 22 GHz \ho masers at component C as seen by VLBI results at mas resolution.  

Analysis of published maser spectra shows that the velocity of the off-nuclear masers has shifted towards the blue during the 35 years covering six observing epochs (Fig. \ref{f8}). 
Since component C results from the shock front between the jet and the ambient molecular materials, this velocity drift towards the observer confirms that the jet points in front of the plane of the sky. 
Thus, the blue-shifted velocity of both the local \hco emission at C and the off-nuclear \ho maser probe the jet-ISM interaction region in the nuclear outflow.

The ring-like structure imaged both at high resolution in the \ho maser and \hco (3--2) molecular gas emission may suggest a shock-driven (cylindrical) expansion of the dense gas triggered by the heating from the decelerating shock in the radio jet. 
The 22\,GHz off-nuclear \ho maser components trace velocity coherent column density regions at the edges of this shock-driven expansion.
 The observed ring-like distribution of the line emission of \hco (3-2) superposed on the maser spots is consistent with the expanding ring model that is recently proposed in \citet{gal23}.

The non-LTE radiative transfer model calculation explains the observed molecular line intensity ratios obtained by the sub-millimeter ALMA observations. 
The results of our calculations using RADEX are largely consistent with conditions where 22\,GHz \ho maser excitation occurs.
Thus, the data obtained from ALMA enabled us to diagnose the gas conditions required for excitation of sub-mm molecular line as well as for the \ho maser emission.

Further observations at higher angular resolution using interferometry with VLBI and ALMA would reveal more details of the gas dynamics and conditions at component C.

\section*{Acknowledgements}
The authors thank T.W.B. Muxlow and A.M.S. Richards for their help in data analysis. YH thanks H.-R.Kl\"{o}ckner and A.Kraus for their observing help at Effelsberg. The authors appreciate H.Sudou for providing a figure used in this article. 
This research was supported by Japan Society for the Promotion of Science (JSPS) Grant-in-Aid for Scientific Research (B) (Y.H.: JP15H03644) and (C) (M.I.: JP21K03632). This research has been supported by the Inoue Enryo Memorial Grant, TOYO University in Japan. 
WAB has been supported by the Chinese Academy of Sciences President's International Fellowship Initiative grants 2022VMA0019 and  2023VMA0030. 
This research has made use of the NASA/IPAC Extragalactic Database (NED), which is funded by the National Aeronautics and Space Administration and operated by the California Institute of Technology.
MERLIN is a National Facility operated by the University of Manchester at Jodrell Bank Observatory on behalf of STFC. 

This article makes use of the following ALMA data: \#2016.1.00052.S, 2016.1.00232.S. ALMA is a partnership of ESO (representing its member states), NSF (USA) and NINS (Japan), together with NRC (Canada) and NSC and ASIAA (Taiwan), in cooperation with the Republic of Chile. 
The Joint ALMA Observatory is operated by ESO, AUI/NRAO and NAOJ. 
Partly based on observations with the 100-m telescope of the MPIfR (Max-Planck-Institut f\"{u}r Radioastronomie) at Effelsberg.  

\section*{Data Availability}
The MERLIN data used in this article will be shared on request.
The molecular line data is available in the publication by \citet{ima20}.
Further data have been obtained from existing literature.









\bsp	
\label{lastpage}
\end{document}